\documentclass[aps,pra,reprint, superscriptaddress]{revtex4-1}

\usepackage{graphicx}
\usepackage{amsmath}
\usepackage{color}
\usepackage{bm}
\usepackage{hyperref}

\newcommand{\TDO}[0]{TiO$_2$}
\newcommand{\epss}[0]{\epsilon}
\newcommand{\kv}[0]{\mathbf{k}}
\newcommand{\rv}[0]{\mathbf{r}}
\newcommand{\la}{\lambda}
\newcommand{\spe}[2]{E^{#1}_{#2}}
\newcommand{\rene}{\tilde{E}}
\newcommand{\ac}[1]{a^\dagger_{#1}}
\newcommand{\ad}[1]{{a}_{#1}}
\newcommand{\qv}{\mathbf{q}}
\newcommand{\pv}{\mathbf{p}}

\newcommand{\eqv}{\hat{\mathbf{e}}_E}
\newcommand{\qEv}{\mathbf{q}_E}
\newcommand{\td}{\frac{\partial}{\partial t}}
\newcommand{\Ev}{\mathbf{E}}
\newcommand{\Bv}{\mathbf{B}}
\newcommand{\Av}{\mathbf{A}}
\newcommand{\UC}{{\Omega_0}}
\newcommand{\Gv}{\mathbf{G}}

\begin{document}

\title{Hybrid cluster-expansion and density-functional-theory approach for
optical absorption in \TDO{}}
\author{O. V\"ansk\"a}
\affiliation{Department of Physics and Material Sciences Center,
Philipps-Universit\"at Marburg, Renthof 5, 35032 Marburg, Germany}
\author{M. Ljungberg}
\affiliation{Department of Physics and Material Sciences Center,
Philipps-Universit\"at Marburg, Renthof 5, 35032 Marburg, Germany}
\affiliation{Donostia International Physics Center, Paseo Manuel
de Lardizabal 4, E-20018 Donostia-San Sebasti\'{a}n, Spain}
\author{P. Springer}
\affiliation{Department of Physics and Material Sciences Center,
Philipps-Universit\"at Marburg, Renthof 5, 35032 Marburg, Germany}
\author{D. S\'anchez-Portal}
\affiliation{Centro de F\'{\i}sica de Materiales CFM-MPC,
Centro Mixto CSIC-UPV/EHU, Paseo Manuel de Lardizabal 5, E-20018
San Sebasti\'an, Spain}
\affiliation{Donostia International Physics Center, Paseo Manuel
de Lardizabal 4, E-20018 Donostia-San Sebasti\'{a}n, Spain}
\author{M. Kira}
\affiliation{Department of Physics and Material Sciences Center,
Philipps-Universit\"at Marburg, Renthof 5, 35032 Marburg, Germany}
\author{S. W. Koch}
\affiliation{Department of Physics and Material Sciences Center,
Philipps-Universit\"at Marburg, Renthof 5, 35032 Marburg, Germany}

\date{\today}

\begin{abstract}
A combined approach of first-principles density-functional calculations and the
systematic cluster-expansion scheme is presented. The dipole, quadrupole, and
Coulomb matrix elements obtained from \emph{ab initio} calculations are used as
an input to the microscopic many-body theory of the excitonic optical response.
To demonstrate the hybrid approach for a nontrivial semiconductor system, the
near-bandgap excitonic optical absorption of rutile \TDO{} is computed.
Comparison with experiments yields strong evidence that the observed
near-bandgap features are due to a dipole-forbidden but quadrupole-allowed
$1s$-exciton state.

\end{abstract}



%71.15.Mb Density functional theory, local density approximation, gradient and
%other corrections
%71.20.Nr Electron density of states and band structure of crystalline solids:
%Semiconductor compounds
%71.35.Cc Excitons and related phenomena: Intrinsic properties of excitons;
%optical absorption spectra
%78.20.Bh Optical properties of bulk materials and thin films: Theory, models,
%and numerical simulation
%78.40.Fy Absorption and reflection spectra: visible and ultraviolet:
%Semiconductors

\maketitle

\section{Introduction}
The single-particle electronic states are the basis needed to quantitatively
model the excitonic optical properties of a semiconductor system
\cite{peter2010fundamentals, haug2009quantum, SQO, mahan2000many}.
In many cases, we can use semiempirical methods like the $\kv\cdot\pv$ theory
together with the effective-mass approximation
\cite{kane_chapter_1966, peter2010fundamentals, haug2009quantum, SQO}
to obtain the electronic wave functions and energies. However, there are many
systems which are not characterized well enough for the needed input parameters
to be reliably known. These kinds of "nontrivial" systems are
found, for example, in novel semiconductor materials containing more than two
constituents, like ternary, quaternary and even more complex compounds
\cite{vurgaftman2001band}, in dilute bismides or nitrides
\cite{broderick2012band, vurgaftman2003band}, in organic systems,
in organic/inorganic heterostructures, and in complex interfaces
\cite{doi:10.1021/cm049654n,graetzel2012materials}.

Even seemingly simple binary systems such as bulk rutile \TDO{} pose
considerable challenges. At first sight, \TDO{} seems to be a common
semiconductor material that has been widely used in applications and intensively
studied over several decades \cite{Fujishima2008515, devore1951refractive,
grant1959properties, pascual1977resolved, amtout1995optical, chiodo2010self,
Migani:2014}. Despite this, after a second look, one quickly realizes that
\TDO{} clearly belongs into the class of nontrivial systems mentioned above,
because many of its essential parameters are poorly known. For example, the
reported conduction-band effective mass in rutile varies almost two orders of
magnitude \cite{pascual1977resolved, madelung2000non, zhang2014new}.
Additionally, it exhibits an exceptionally large refractive index in the visible
range as well as particularly strong birefringence and dispersion properties
\cite{devore1951refractive, grant1959properties}.

Furthermore, rutile has highly asymmetric dielectric properties regarding
different crystallographic directions with a particularly large magnitude of the
low-frequency dielectric constant \cite{madelung2000non}. This results in an
exceptionally strong screening of the Coulomb interaction between electrons and
holes for excitons with a binding energy below or comparable to the polar phonon
energies \cite{bechstedt2015many}. Nevertheless, optical absorption measurements
\cite{pascual1977resolved, pascual1978fine, amtout1992time, amtout1992resonant,
amtout1995optical} near the band gap of \TDO{} have shown some indications of
excitonic features which, however, are very controversially discussed in the
literature. For example, the excitonic signatures were interpreted as a
dipole-forbidden but quadrupole-allowed $1s$ exciton \cite{pascual1977resolved,
pascual1978fine}, or as a weakly dipole-allowed $2p$ exciton state
\cite{amtout1992time, amtout1992resonant, amtout1995optical}.

In order to study the optical properties of rutile and to develop a general
scheme that allows us to overcome the restrictions of the semiempirical models,
we have to use more systematic methods to access the microscopic properties of
the electronic states. Here, the most widely used scheme is density functional
theory (DFT)  \cite{HohenbergKohn:1964, KohnSham:1965}, which has been proven to
be a very efficient approach in obtaining ground-state properties for various
solid-state systems, molecules, nanostructures, liquids, and molecules adsorbed
on surfaces \cite{Martin:2004, Eberhard:2011}.

The DFT computed single-particle energies and wave functions can then be used as
an input to the cluster-expansion scheme \cite{PQE06, SQO} that provides
first-principles level description of many-body dynamics. Its strength relies on
a systematic grouping of different correlations as clusters that build up
sequentially in time \cite{SQO}. The cluster-expansion method is widely applied
in semiconductor solid-state systems \cite{richter2009few, PhysRevB.83.205305,
harsij2012influence, leymann2013intensity, almand2014quantum, leymann2015sub,
suwa2014coherent} and it even can describe quantum-optical properties
\cite{kira2006quantum, *PhysRevA.78.022102} as well as strongly interacting Bose
gas \cite{Kira2014200, *kira2015hyperbolic, *kira2015coherent} quantitatively.

In this work, we combine a DFT approach with the cluster-expansion method to
obtain a hybrid cluster-expansion and density-functional-theory (CE-DFT) scheme
for studying dynamical many-body quantum phenomena. We first obtain the needed
Coulomb and optical matrix elements via DFT for the second-quantization system
Hamiltonian that we then use to model dynamical effects by applying the
cluster-expansion scheme. The electronic excitation dynamics are derived in a
very general form, yielding a structure similar to the semiconductor Bloch
equations \cite{lindberg1988effective, haug2009quantum}. Prior to our work,
\emph{ab initio} matrix elements have been used together with the semiconductor
Bloch equations to compute the surface exciton properties on a Si surface 
\cite{PhysRevB.68.045330}. Linear optical properties such as the absorption
follow directly once the dynamics have been solved. We then apply our combined
CE-DFT method for rutile \TDO{} and study its near-bandgap optical properties by
considering electric-dipole, electric-quadrupole, and magnetic-dipole
light--matter interactions. We also show that the electric-quadrupole
interaction is highly dependent on the propagation and polarization directions
of the light. We use our microscopic results to analyze the experimentally
available absorption spectra \cite{pascual1977resolved, pascual1978fine,
amtout1992time, amtout1992resonant, amtout1995optical}, finding a good level of
agreement that allows us to identify the observed excitonic signature as a
dipole-forbidden but quadrupole-allowed $1s$ exciton.

%%%%%%%%%%%%%%%%%%%%%%%%%
%%%%%%%%%%%%%%%%%%%%%%%%%
\section{Theoretical Background}
To effectively formulate the quantum kinetics of a semiconductor system via the
cluster-expansion method, we start from a generic Hamiltonian \cite{SQO, PQE06}
\begin{align}
\label{sysH}
H=&\sum_{i} E_i \,\ac{i} \ad{i} -E(t) \sum_{i,j}\, F_{ij}\, \ac{i} \ad{j} 
\nonumber \\
& + \frac{1}{2}\sum_{i, i',j, j'} V_{ii';j'j} \, \ac{i} \ac{i'} \ad{j'}\ad{j},
\end{align}
where $\ac{i}$ ($\ad{i}$) is the creation (annihilation) operator of an
electronic state defined by the particle index $i$ and $E(t)$ is the
time-dependent electric field. This Hamiltonian is uniquely defined by
$E_i$, $F_{ij}$, and $V_{ii';j'j}$, i.e., by the energies of the electronic
states and the matrix elements for light--matter and electron--electron
interactions, respectively.

To identify these input parameters from a DFT calculation, we repeat the
derivation of Eq.~(\ref{sysH}) from the Lagrangian and Hamiltonian formalism of
classical electrodynamics \cite{cohen1997photons}. In the Coulomb gauge, the
quantization yields the conventional minimal-substitution
Hamiltonian for electronic quasiparticles \cite{SQO, kira1999quantum}:
\begin{align}
\label{H_min}
H_N=&\sum_{i=1}^N\bigg\{\frac{1}{2m_0}\left[\pv_i-e\Av(\rv_i,t)\right]^2
+ U(\rv_i)\bigg\}
\nonumber \\
&\quad + \frac{1}{2}\sum_{i\neq j} V(\rv_i,\rv_j),
\end{align}
where $m_0$ is the free-electron mass, $e$ is the charge of an electron,
$\Av(\rv,t)$ is the vector potential of an optical field, $U(\rv)$ is an
external potential, and $V(\rv,\rv')$ is a pairwise electron--electron
interaction.

Equation~(\ref{H_min}) includes a contribution proportional to the square of the
optical field that can be eliminated via the Power-Zienau-Woolley transformation
\cite{cohen1997photons, keller2011quantum} in our initial Coulomb-gauge
Lagrangian. This way, we obtain a classical Hamiltonian where the light--matter
interaction is directly given by the electric $\Ev(\rv,t)$ and the magnetic
$\Bv(\rv,t)$ fields instead of the vector potential. After quantization, it
reads
\begin{align}
\label{H_min_2}
H'_N= & \sum_{i=1}^N H(\pv_i,\rv_i) - E(t)\sum_{j=1}^N F(\pv_i,\rv_i)
\nonumber \\
&+\frac{1}{2}\sum_{i\neq j} V(\rv_i,\rv_j) +H_{\text{RM}}, 
\end{align}
where we assumed fields of the form $\Ev(t,\rv)=E(t)\Ev(\rv)$ and
$\Bv(t,\rv)=E(t)\Bv(\rv)/\omega$ with the optical frequency $\omega$. The
diamagnetic contributions \cite{cohen1997photons, keller2011quantum}
$H_{\text{RM}}$ can be neglected at moderate field intensities, hence we omit
this term in the following. The single-particle Hamiltonian is given by
\begin{equation}
\label{spH}
	H(\pv,\rv)=\frac{\pv^2}{2m_0} + U(\rv),
\end{equation}
and the light--matter-interaction operator by 
\begin{align}
\label{MPoper}
&F(\pv,\rv) =  -e \rv \cdot \sum_{n=0}^{\infty}\frac{1}{(n+1)!}
[(\rv \cdot\nabla_{\rv'})^n \, \Ev(\rv')]_{\rv'=0} 
\nonumber \\
\ &-\frac{e}{2 m_0} \sum_{l=0}^{\infty} \frac{l+1}{(l+2)!} \Big\{ \pv \cdot
\Big[\rv \times [(\rv \cdot \nabla_{\rv'})^l \, \Bv(\rv')]\Big]
\nonumber \\
\ &+\Big[\rv \times [(\rv \cdot\nabla_{\rv'})^l \, \Bv(\rv')]\Big]
\cdot\pv \Big\}_{\rv'=0},
\end{align}
after expressing $\Ev(\rv)$ and $\Bv(\rv)$ through their multipole expansions.
The electric-dipole interaction ($n=0$) can be weak, especially near the
direct band gap of \TDO{} \cite{pascual1977resolved, pascual1978fine}. To keep
our approach general while simultaneously suitable for near-bandgap optical
properties of \TDO{}, we initially consider all terms within the multipole
expansions of $\Ev(\rv)$ and $\Bv(\rv)$.

If we know the electronic wave functions $\phi_i(\rv)$ and energies $E_i$ that
follow from the Schr\"odinger equation
\begin{equation}
\label{SPSC}
	H(\pv,\rv) \, \phi_i(\rv) = E_i \, \phi_i(\rv),
\end{equation}
we can construct the fermionic field operator
\begin{equation}
\label{FO}
	\hat{\Psi}(\rv)=\sum_{i} \phi_{i}(\rv) \, \ad{i}.
\end{equation}
As elaborated in many text books \cite{mahan2000many, SQO, haug2009quantum}, we
can use $\hat{\Psi}(\rv)$ in the second-quantization step for $H(\pv,\rv)$,
$F(\pv,\rv)$, and $V(\rv,\rv')$ to obtain the Hamiltonian~(\ref{sysH}) from
Eq.~(\ref{H_min_2}) with the matrix elements
\begin{align}
\label{GDME}
F_{ij} = & \int d^3 r \, \phi^*_{i}(\rv) \, F(\pv,\rv)\,  \phi^*_{j}(\rv), \\
\label{GCME}
V_{ii';j'j} = & \int d^3 r  d^3 r'\, \phi^*_{i}(\rv) \phi^*_{i'}(\rv') 
V(\rv,\rv')  \phi_{j'}(\rv') \phi_{j}(\rv).
\end{align}

In principle, $\phi_i(\rv)$ could be chosen to be any complete orthonormal set
for the second-quantization step. In practice, it is useful if they are
eigenfunctions of a suitable effective Hamiltonian that is known to give
eigenvalues $E_i$ corresponding to accurate ionization energies and electron
affinities. A possible choice is a DFT Hamiltonian with a scissor shift to
obtain the correct band gap, even better would be to use a scissor shifted
static $GW$ Hamiltonian if it is available (we avoid explicit frequency
dependence in the single-particle Hamiltonian in order to keep the time
evolution simple and ensure orthogonal wave functions). The unshifted
Hamiltonian can be written by replacing the potential $U(\rv)$ with
\begin{equation}
U[\rho_0](\rv) = V_{\text{ext}}(\rv) + V_{\text{H}}[\rho_0](\rv)
+ V_{\text{xc}}[\rho_0](\rv), 
\end{equation}
where the external potential from the nuclei $V_{\text{ext}}(\rv)$ is augmented
with an effective potential that includes the interaction of the electrons with
the average electron density $V_{\text{H}}[\rho_0](\rv)$, as well as approximate
exchange and correlation effects $V_{\text{xc}}[\rho_0](\rv)$. Note that these
terms are explicitly dependent on the equilibrium density matrix $\rho_0$ that
is determined in DFT by a self-consistent iterative procedure. To correct the
band gap we shift the eigenvalues by a downshift for the states below the Fermi
level and an upshift on the states above it. The terms we have added to $U(\rv)$
include some Coulomb terms that must be later subtracted when we develop the
equations of motion as we discuss in Appendix~\ref{RQ}.

One of the most critical conditions for the success of our CE-DFT method is that
the computed $\phi_{i}(\rv)$ and $E_i$ are sufficiently close to the actual
one-particle wave function and energies, respectively. When we assume that this
condition is fulfilled, by utilizing Eqs.~(\ref{GDME}) and (\ref{GCME}), we
obtain a DFT-defined system Hamiltonian~(\ref{sysH}) that is the basis of our
hybrid CE-DFT approach.

%%%%%%%%%%%%%%%%%%%%%%%%%
%%%%%%%%%%%%%%%%%%%%%%%%%
\section{CE-DFT equation for optical absorption}
\label{sec_Abso}
It is beneficial to divide the electronic index $i$ in Eq.~(\ref{sysH}) into two
parts $i\rightarrow (\la,j)$ where $\la=v$ or $\la=c$ for states that are either
occupied or unoccupied in the ground state of the system, respectively, and $j$
defines the exact state inside these groups. After we have constructed our
DFT-defined system Hamiltonian $H$, we can study the dynamics of the expectation
values $P^{\la\la'}_{ij}\equiv \langle \ac{\la,i}\ad{\la',j} \rangle$ via the
Heisenberg equation of motion and the cluster-expansion method
\cite{PQE06, SQO}. This yields the semiconductor Bloch equations
\begin{align}
\label{SBE}
i\hbar  \td  P^{\la\la'}_{ij} = &  \sum_k \big[ \rene^{\la'}_{kj} \, 
P^{\la\la'}_{ik} - \rene^{\la}_{ik} \, P^{\la\la'}_{kj} + \Omega^{\la}_{ik} \,
P^{\bar{\la}\la'}_{kj}
\nonumber \\
&-\Omega^{\bar{\la}'}_{kj}\,P^{\la \bar{\la}'}_{ik} \big]+\Gamma^{\la\la'}_{ij},
\end{align}
where $\rene^{\la}_{ij}$ ($\Omega^{\la}_{ij}$) is the renormalized kinetic
energy (Rabi frequency) and $\bar{\la}$ denotes the complement of $\la$
($\bar{v}=c$ and $\bar{c}=v$). Explicit forms of $\rene^{\la}_{ij}$ and
$\Omega^{\la}_{ij}$ can be found in Appendix~\ref{RQ}. The term 
$\Gamma^{\la\la'}_{ij}$ includes the coupling to the higher-order clusters that
follows from the two-particle contributions \cite{PQE06, SQO} due to Coulomb and
phonon interactions.

Similarly to the formulation of Eq.~(\ref{SBE}), we can construct equations for
higher order clusters and study, e.g., dynamics of exciton
correlations \cite{PQE06, SQO}. However, in this work we focus on fundamental
aspects of combining DFT and the cluster-expansion method, thus limiting our
considerations on the level of Eq.~(\ref{SBE}).

In this situation, Eq.~(\ref{SBE}) yields a closed set of equations between the
dynamics of the microscopic polarization $P_{ij}\equiv P^{vc}_{ij}$, the
electron occupation $f^{e}_{ij}\equiv P^{cc}_{ij}$, and the hole occupation
$f^{h}_{ij} \equiv \delta_{ij}-P^{vv}_{ij}$. Even though this set of equations
can be used to model many interesting dynamical effects such as nonlinearities
or excitation induced changes in the optical response, in the present study we
concentrate on the linear absorption properties of the rutile \TDO{} system.

In this case, $f^{e}_{ij}=f^h_{ij}=0$ in Eq.~(\ref{SBE}) and we can neglect
several terms that are connected to optical nonlinearities. When we also make
the Tamm-Dancoff approximation \cite{Onida_Reining:2002} by omitting processes
that would generate coupling between the $P^{vc}_{ij}$ and $P^{cv}_{ij}$ terms,
Eq.~(\ref{SBE}) reduces to (see Appendix~\ref{RQ} for details)
\begin{align}
\label{PSBE}
i\hbar  \td  P_{ij} = & [\spe{c}{j}-\spe{v}{i}]P_{ij}-\sum_{k,l}
{\mathcal{V}}^{vc;cv}_{kj;li} \, P_{kl}
\nonumber \\
&-E(t) \, F^{cv}_{ji} +\Gamma^{vc}_{ij},
\end{align}
containing
\begin{equation}
\label{CandE}
{\mathcal{V}}^{\la_1\la_2;\la_3 \la_4}_{i_1i_2;i_3i_4}\equiv
V^{\la_1\la_2;\la_3 \la_4}_{i_1i_2;i_3i_4}
- V^{\la_2\la_1;\la_3 \la_4}_{i_2i_1;i_3i_4},
\end{equation}
where we have grouped the direct and exchange electron--electron interactions
under a single matrix element. Here, we use a notation where the $\la$ index in
the matrix elements is written as a superscript above the corresponding $i$
index, e.g., $\spe{v}{i}\equiv E_{v,i}$. While Eq.~(\ref{SBE}) can be used for
any system defined by the Hamiltonian~(\ref{H_min}) and when rotating-wave
contributions dominate the dynamics, Eq.~(\ref{PSBE}) is applicable only in
cases where the Tamm-Dancoff approximation is justified.

The homogeneous solution of Eq.~(\ref{PSBE}) defines the Wannier equation
\cite{SQO, haug2009quantum}
\begin{align}
\label{wannier}
[\spe{c}{j}-\spe{v}{i}]\phi_{\nu}(i,j)&-\sum_{k,l} {\mathcal{V}}^{vc;cv}_{kj;li}
\, \phi_{\nu}(k,l)
\nonumber \\
&= E_{\nu} \, \phi_{\nu}(i,j),
\end{align}
as an eigenvalue problem for the wave function $\phi_{\nu}(i,j)$ and the energy
$E_{\nu}$ of an exciton state $\nu$. Converting the polarization into the
exciton basis,
\begin{equation}
\label{Xbasis}
	P_{ij} = \sum_\nu p_\nu \, \phi_{\nu} (i,j),\quad	
	p_\nu = \sum_{i,j} \phi^*_{\nu} (i,j) \, P_{ij},
\end{equation}
the Fourier transform of Eq.~(\ref{PSBE}) yields
\begin{equation}
\label{SBE2}
[\hbar\omega +i \gamma_\nu(\omega)] p_\nu(\omega)=E_\nu \, p_\nu(\omega)
-{\mathcal{F}}_\nu \, E(\omega),
\end{equation}
where $E(\omega)$ is the Fourier transform of $E(t)$ and
\begin{equation}
\label{oss}
	{\mathcal{F}}_\nu\equiv \sum_{i,j} \phi^*_{\nu} (i,j) \, F^{cv}_{ji}
\end{equation}
defines a generalized oscillator strength of optical transitions.

To avoid a detailed analysis of Coulomb and phonon scattering, we introduce a
dephasing function 
\begin{equation}
\label{dephasing}
\gamma_\la(\nu)=\frac{\gamma_\nu}
{\exp[(E_\nu-\hbar\omega-\mu_\nu)/\Delta E_\nu] +1 }
\end{equation}
that phenomenologically includes the contributions of the
$\Gamma^{vc}_{ij}$ term in Eq.~(\ref{SBE2}). With the parameters $\gamma_\nu$,
$\mu_\nu$, and $\Delta E_\nu$, function $\gamma_\nu(\omega)$ describes the
excitation-induced dephasing effects \cite{PQE06, smith2010extraction} as well
as the exponential decay of the absorption tail towards lower energies;
a phenomenon known as the \emph{Urbach tail}
\cite{haug2009quantum, kurik1971urbach}, which is of particularly significance
in polar semiconductors \cite{liebler1991calculation} like \TDO{}.

We can now solve Eq.~(\ref{SBE2}) directly, yielding
\begin{equation}
\label{pol}
p_\nu(\omega)=\frac{{\mathcal{F}}_\nu}{E_\nu-\hbar\omega-i \gamma_\nu(\omega)}
E(\omega).
\end{equation}	
The absorption response of our system then follows from the imaginary part of
the linear susceptibility \cite{SQO}
$\chi(\omega)=\frac{P(\omega)}{\epsilon_0 E(\omega)}$ where
$P(\omega)=\sum_\nu {\mathcal{F}}_\nu^* \, p_\nu(\omega)$ is the macroscopic
polarization. With the help of Eq.~(\ref{pol}), the Elliott formula for the
optical absorption becomes
\begin{align}
\label{abso}
\alpha(\omega) & \equiv \frac{\omega}{n c}\text{Im}[\chi(\omega)]
\nonumber \\
&= \frac{\omega}{\epsilon_0  n c}\sum_{\nu} \frac{|{\mathcal{F}}_\nu|^2 \,
\gamma_\nu(\omega) }{(E_\nu-\hbar\omega)^2+\gamma^2_\nu(\omega)},
\end{align}
where $n$ is the refractive index of the material.

%%%%%%%%%%%%%%%%%%%%%%%%%
%%%%%%%%%%%%%%%%%%%%%%%%%
\section{Excitonic absorption in rutile}
\subsection{Two-band model}
\label{sec_EAtio}
To apply the general results of the previous section for \TDO{}, we assume a
three-dimensional crystal having an infinite volume. Then, the potential
$U(\rv)$ in the Hamiltonian~(\ref{spH}) must have the periodicity of the
lattice, and we can use the Bloch theorem to produce electronic wave
functions of the form
\begin{equation}
\label{SPWF}
\phi_{\la,\kv}(\rv) = \frac{e^{i\kv\cdot\rv}}{(2\pi)^{3/2}}\,u_{\la,\kv}(\rv),
\end{equation}
where $u_{\la,\kv}(\rv)$ is a lattice-periodic Bloch function, $\kv$ is the wave
vector, and $\la$ denotes different Bloch bands. Thus, in the main step before
constructing the system Hamiltonian~(\ref{sysH}) of our crystal, we solve the
band structure $\spe{\la}{\kv}$ and the Bloch functions $u_{\la,\kv}(\rv)$
corresponding to the Schr\"odinger equation~(\ref{SPSC}) via DFT, requiring the
form of Eq.~(\ref{SPWF}).

Our DFT results presented in Sec.~\ref{BS} reveal that it is well justified to
model the near-bandgap optical properties of \TDO{} by including only the
energetically lowest conduction band $\la=c$ and the highest valence band
$\la=v$. Hence, we can relate the Bloch-band index $\la$ and the $\la$ index of
Sec.~\ref{sec_Abso} and directly use Eqs.~(\ref{SBE})-(\ref{oss}) if we replace
the sums over the particle index by $k$-space integrals. Furthermore, in the
vicinity of the band gap $E_g$ we find (see Sec.~\ref{BS}) that the electron
energy $\spe{e}{\kv} \equiv \spe{c}{\kv}-E_g$ and the hole energy
$\spe{h}{\kv} \equiv -\spe{v}{\kv}$ are accurately described by a parabolic
dispersion
\begin{equation}
\label{parabolic}
	\spe{\la}{\kv}=\frac{\hbar^2}{2 m^{\la}_x}k_x^2 
	+\frac{\hbar^2}{2 m^{\la}_y}k_y^2+\frac{\hbar^2}{2 m^{\la}_z}k_z^2,
\end{equation}
where $m^{\la}_{i}$ is the $i$-directional effective mass of an electron or hole
for $\la=e$ or $\la=h$, respectively.

%%%%%%%%%%%%%%%%%%%%%%%%%
\subsection{Light--matter interaction}
When we express $\Ev(\rv)$ and $\Bv(\rv)$ in Eq.~(\ref{MPoper}) via their
multipole expansions, the matrix element in Eq.~(\ref{GDME}) is related to the
position operator $\rv$. In our crystal system, the position operator $\rv$ maps
a wave function by $\rv\phi_{\la,\kv}(\rv)$ out of the Hilbert space that is
constructed by the lattice-periodic wave functions \cite{QUA:QUA25}.
Consequently, the $\rv$ matrix element is not uniquely defined and, for example,
in DFT computations its value depends on how we chose the unit cell of a system
\cite{gu2013relation}. This is a known problem that has been widely studied in
connection with the electric-dipole matrix element over several decades
\cite{yafet1957g, blount1962formalisms, resta1994macroscopic, gu2013relation}
and is still an important research topic \cite{swiecicki2014linear}.

In general, while the light--matter matrix element in Eq.~(\ref{GDME}) is not
directly observable, it manifests itself through other quantities like the
oscillator strength ${\mathcal{F}}_\nu$. By considering 
$F^{\la\la'}_{\kv\kv'}$ in Eq.~(\ref{oss}) for an infinite crystal under the
assumption that a two-band model with a parabolic band structure is valid,
we present in Appendix~\ref{LM} a method where the ambiguities related to the
$\rv$ matrix elements can be avoided by expressing the light--matter matrix
elements via the momentum matrix element $\pv^{\la\la'}_{\kv}$. Using our DFT
Bloch functions, these matrix elements are computed via
\begin{equation}
\label{pelem}
\pv^{\la\la'}_{\kv} \equiv \frac{1}{\Omega_0} \int_{\Omega_0} d^3 r\,
u^*_{\la,\kv}(\rv) \, \pv \, u_{\la',\kv}(\rv),
\end{equation} 
with the unit cell volume $\Omega_0$.

Even though we discuss in Appendix~\ref{LM} how our approach can be used to
solve $F^{cv}_{\kv\kv'}$ for arbitrary $\Ev(\rv,t)$ and $\Bv(\rv,t)$, we focus
on the electric-dipole, electric-quadrupole, and magnetic-dipole light--matter
interactions. Furthermore, we assume that a plane wave with
$\mathbf{E}(\rv)=e^{i\qEv\cdot\rv}\eqv$ and
$\mathbf{B}(\rv)=e^{i\qEv\cdot\rv}(\qEv\times\eqv)$ propagates through our
sample, where $\qEv$ and $\eqv$ define the wave-vector and the polarization
direction of the optical field, respectively. Then, the $n=0$, $n=1$, and $l=0$
terms of Eq.~(\ref{MPoper}) yield the light--matter interaction operators
\begin{align}
\label{Dope}
	D \equiv &-e \eqv \cdot \rv,  \\ 
\label{Qope}
	Q \equiv & -i \frac{e}{2} (\eqv \cdot \rv)(\qEv \cdot \rv),  \\
\label{Mope}
	M \equiv & -\frac{e}{2\omega m_0}[ \rv \times (\qEv \times \eqv)]\cdot  \pv,
\end{align}
for the electric-dipole, electric-quadrupole and magnetic-dipole interactions,
respectively.

By using the steps outlined in Appendix~\ref{LM}, we connect the operators
$D$, $Q$, and $M$ to the matrix elements
\begin{align}
\label{DME}
D^{cv}_{\kv\kv'} = &\delta(\kv-\kv') D^{cv}_{\kv}, \\
\label{QME}
Q^{cv}_{\kv\kv'} = &\delta(\kv-\kv') \sum_{i,j} e_i\, Q^{cv}_{\kv;ij}q_j,\\
\label{MME}
M^{cv}_{\kv\kv'} = &\delta(\kv-\kv') \sum_{i,j} e_i \, M^{cv}_{\kv;ij}q_j.
\end{align}
Here, $e_i$ and $q_i$ are the $i$-directional components of the vectors
$\eqv$ and $\qEv$, respectively,
${D}^{cv}_{\kv} = -i e  \eqv \cdot \tilde{\pv}^{cv}_\kv$ includes the
conventional \cite{haug2009quantum, SQO, gu2013relation} relation between the
$\rv$ and $\pv$ matrix elements via
\begin{equation}
\label{pmat}
\tilde{\pv}^{\la\la'}_\kv \equiv \frac{\hbar}{m_0}\frac{\pv^{\la\la'}_{\kv}}
{\spe{\la}{\kv}-\spe{\la'}{\kv}},
\end{equation}
and the explicit forms of ${Q}^{cv}_{\kv;ij}$ and ${M}^{cv}_{\kv;ij}$ are given
in Appendix~\ref{LM}. The matrix element $F^{cv}_{\kv\kv'}$ becomes
\begin{equation}
\label{Fele_final}
	F^{cv}_{\kv\kv'}= \delta(\kv-\kv')(D^{cv}_{\kv}+Q^{cv}_{\kv}+M^{cv}_{\kv}).
\end{equation}

%%%%%%%%%%%%%%%%%%%%%%%%%
\subsection{Coulomb interaction and direct excitonic states}
For rutile \TDO{}, it is necessary to model the electron--electron interaction
$V(\rv,\rv')$ using the result for anisotropic media
\cite{landau1984electrodynamics}. The resulting expression can be given through
the Fourier transformation
\begin{align}
\label{CI}
	V(\rv,\rv') = &\int d^3 k\,  V_{\kv} e^{i\kv\cdot(\rv-\rv')}, 	 \\
\label{CME}
	V_{\kv} = & \frac{e^2}{8\pi^3\varepsilon_0}\frac{1}{\epsilon_x k_x^2
	+\epsilon_y k_z^2+\epsilon_z k_z^2}, 
\end{align}
where $\varepsilon_0$ is the vacuum permittivity, $\epsilon_i$ is the dielectric
constant along the $i$-directional principal axis of the permittivity tensor
\cite{landau1984electrodynamics} and $k_i$ is the related component of the $\kv$
vector, respectively. If our system has an excitation in a sufficiently small
region of the Brillouin zone, we can quite generally approximate the
Coulomb matrix element by
\begin{equation}
\label{SC_CME}
	V^{\la_1 \la_2; \la_3 \la_4}_{\kv_1 \kv_2;\kv_3 \kv_4}
	=\delta(\kv_1+\kv_2-\kv_3-\kv_4)\delta_{\la_1\la_4}
	\delta_{\la_2\la_3}V_{\kv_1-\kv_4},
\end{equation}
which follows from Eqs.~(\ref{GCME}), (\ref{CI}), and (\ref{CME})
(see Appendix~\ref{Cint} and Ref.~\cite{bechstedt2015many}).
By using $u_{c,\kv}(\rv)$ and $u_{v,\kv}(\rv)$ in \TDO{} determined by DFT
computations, we carefully check in Appendix~\ref{Cint} that the relevant
Coulomb interaction is indeed well approximated by the matrix 
element~(\ref{SC_CME}).

Since the light--matter and Coulomb interactions conserve the crystal momentum
$\hbar\kv$ via the matrix elements in Eqs.~(\ref{Fele_final}) and
(\ref{SC_CME}), we find from Eq.~(\ref{SBE}) that only direct quantities
$P^{\la\la'}_{\kv}\equiv P^{\la\la'}_{\kv\kv}$ couple to optical excitation if
we omit possible nondiagonalities in $\Gamma^{\la\la'}_{\kv\kv'}$. Consequently,
we need to consider only the direct excitonic states with wave functions
$\phi_{\nu}(\kv) \equiv \phi_{\nu}(\kv, \kv)$ for the optical response following
form Eq.~(\ref{Xbasis}). These states are solutions of the Wannier equation
\begin{equation}
\label{wannier_rel}
	(\spe{c}{\kv}-\spe{v}{\kv})\phi_{\nu}(\kv) -\int d^3 k' \, V_{\kv-\kv'}
	\,\phi_{\nu}(\kv') = E_{\nu} \, \phi_{\nu}(\kv),
\end{equation}
where $\kv$ can be associated with the relative motion of electrons and holes,
and we have assumed that the approximations in Eqs.~(\ref{parabolic}) and
(\ref{SC_CME}) are valid \cite{SQO,bechstedt2015many}.
In Eq.~(\ref{wannier_rel}), the $\nu$ index typically refers to the symmetry of
the exciton states with values $\nu=1s$, $2p$, etc. In an anisotropic system, we
can assume that this symmetry based grouping is only approximate and the
character of, e.g., $s$ and $p$ like states becomes mixed.

Equation~(\ref{wannier_rel}) can have simultaneously anisotropic 
energies $\spe{\la}{\kv}$ and matrix elements $V_{\kv}$, which may complicate
solution strategies. However, we can always perform a $k$-space coordinate
transform in Eq.~(\ref{wannier_rel}) that results in a transformed spherical
Coulomb matrix element (see Appendix~\ref{Xs}). The use of
$V^{\la_1 \la_2; \la_3 \la_4}_{\kv_1 \kv_2;\kv_3 \kv_4}$ in Eq.~(\ref{SC_CME})
together with this coordinate transformation comprises a rather general and
remarkably beneficial approach for modeling properties of anisotropic systems.
This is due to the fact that DFT can be effectively used to compute
$\spe{\la}{\kv}$ and $F^{\la\la'}_{\kv\kv'}$ whereas computing all the
Coulomb-interaction terms would be a numerically highly demanding task;
especially in three-dimensional problems. Furthermore, a spherical
Coulomb matrix element makes it possible to solve the excitonic states and the
related properties of a system efficiently by using techniques developed
elsewhere \cite{PS_unpublished}. In particular, we expand the excitonic wave
functions $\phi_{\nu}(\kv)$ and all matrix elements in terms of spherical
harmonics in the transformed coordinate system and then solve the
corresponding high-dimensional eigenvalue problem given by
Eq.~(\ref{wannier_rel}) as discussed in Appendix~\ref{Xs}
and in Ref.~\cite{PS_unpublished}.

%%%%%%%%%%%%%%%%%%%%%%%%%
%%%%%%%%%%%%%%%%%%%%%%%%%
\subsection{Band structure}
\label{BS}
\begin{figure}
	\includegraphics[width=86mm]{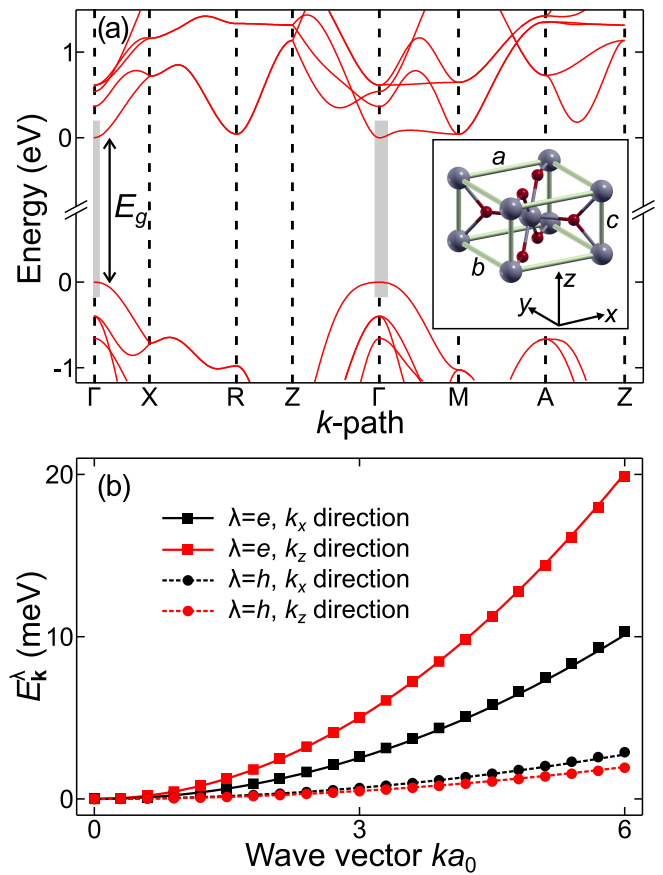}
	\caption{\label{fig1}
	(Color online) \emph{Ab initio} band structure and its parabolic
	approximation. 	(a) A wide region of the \emph{ab initio} band structure
	throughout the whole Brillouin zone, where the $k$-space region of interest
	is indicated by the shaded areas. (b) Electron (solid lines) and hole
	(dashed lines) energies for $k_y=0$ in the region of interest as a function
	of $k_x$ (black lines) and $k_z$ (red lines). Parabolic fits for electron and
	hole energies (lines) approximate \emph{ab initio} energies
	(squares and circles) to a high degree of accuracy.
	}
\end{figure}

In our numerical evaluations, we first set-up our computational cell based on
the experimentally determined \cite{madelung2000non} tetragonal crystal
structure of rutile \TDO{} that is defined by the unit cell with lattice
constants $a=b=4.59$~\AA{} and $c=2.96$~\AA{}. The inset of Fig.~\ref{fig1}(a)
depicts the unit cell in a Cartesian coordinate system where oxygen (titanium)
atoms are illustrated by the red (gray) spheres.

Starting from this unit cell, we perform a DFT calculation with the SIESTA
\cite{SIESTA} code using the PBE exchange-correlation functional
\cite{PBE:1996}, Trouiller-Martins pseudopotentials
\cite{TrouillerMartins:1991}, and a DZP basis set of atomic-like orbitals with
an energy shift parameter of 100~meV. After a sufficient convergence of the
ground-state properties (with 11x11x17 $k$ points), we rediagonalize the
Kohn-Sham Hamiltonian matrix at the $k$ points we are interested in, obtaining
the needed $\spe{\la}{\kv}$ and $u_{\la,\kv}(\rv)$.

To compute the matrix elements in Eqs.~(\ref{GDME}) and (\ref{GCME}), we put the
cell-periodic part of the wave functions on a real space grid and Fourier
transform them to obtain their plane-wave representation. All needed matrix
elements can then be computed either in real or reciprocal space. We use the
DFT computed energies to obtain the effective masses, while the
light--matter-interaction matrix elements in Eq.~(\ref{GDME}) are used directly.

Our computed Coulomb matrix elements in Eq.~(\ref{GCME}) were found to
correspond well to the matrix element~(\ref{SC_CME}) for the small
$(\kv_1-\kv_4)$-values that are important for excitons and we thus chose to use
the latter for simplicity. In addition of confirming the
approximation~(\ref{SC_CME}), the \emph{ab initio} Coulomb matrix elements were
used in a phase matching procedure to obtain the
light--matter-interaction matrix elements that correspond to the phase fixing
we do when we adopt the matrix element~(\ref{SC_CME}). Following this procedure,
we obtain all needed matrix elements for Eq.~(\ref{sysH}), which
establishes the technical basis of our CE-DFT approach in \TDO{}.
%%%%%%%%%%%%%%%%%%%%%%%%%%%%%%%%%%%%%%%%%%%%%%%%%%
%%%%%%%%%%%%%%%%%%%%%%%%%%%%%%%%%%%%%%%%%%%%%%%%%%

The actual $\spe{\la}{\kv}$ dispersion obtained from our DFT computations is
shown in Fig.~\ref{fig1}(a) along the high-symmetry points of the
Brillouin zone. The depicted band structure agrees to some extent with multiple
previous results
\cite{PhysRevB.46.1284, mo1995electronic, landmann2012electronic} and indicates
a direct gap at the $\Gamma$ point. Our DFT approach with local functionals
underestimates the band gap $E_g$ by 1.3~eV, which is corrected with the
well-known ``scissor shift'' to produce the experimental $E_g=3.03$~eV
\cite{pascual1977resolved, madelung2000non}. However, we use band energies
$\spe{\la}{\kv}$ given directly by our DFT computations in Eq.~(\ref{pmat})
that is linked to all light--matter-interaction matrix elements.

The shaded rectangles in Fig.~\ref{fig1}(a) indicate the electronic states most
important for near-bandgap optical transitions, covering roughly a range of
20~meV in electron--hole energy, $\spe{c}{\kv}-\spe{v}{\kv}$, around the
$\Gamma$ point. In this region, the energetically lowest conduction and highest
valence band are relatively well separated from the nearby bands such that we 
can adopt the two-band approach described in Sec.~\ref{sec_EAtio}.
Figure~\ref{fig1}(b) shows the electron (continuous lines and squares) and the
hole energies (dashed lines and circles) near the $\Gamma$ point as a function
of $k_x$ (black lines) and $k_z$ (red lines) when $k_y=0$.
The reciprocal $k_x$, $k_y$, and $k_z$ directions correspond to lattice
$x$, $y$, and $z$ directions, respectively. These directions are selected
to be parallel to the $a$, $b$, and $c$ axes of \TDO{} crystal structure,
as indicated in the insert of Fig.~\ref{fig1}(a).

The computed \emph{ab initio} $\spe{e}{\kv}$ and $\spe{h}{\kv}$ energies are
almost parabolic in all directions and symmetric with respect to
$k_x \leftrightarrow k_y$ exchange. However, $k_x$ (or $k_y$) and $k_z$
directions exhibit a different $k$ dependence, which creates a strong
anisotropy. This anisotropy can be accurately described by
Eq.~(\ref{parabolic}) with effective masses $m^e_x=m^e_y=1.03m_0$,
$m^e_z=0.519m_0$, $m^h_x=m^h_y=3.80m_0$, and $m^h_z=5.30m_0$.
Figure~\ref{fig1}(b) compares the parabolic model (lines) and the
\emph{ab initio} energies (squares and circles), showing the high accuracy of
our approximation. In this figure, the wave vector is scaled by the exciton
Bohr radius $a_0\equiv4\pi\epsilon_0\epss\hbar^2/(e^2\mu)$ with geometrically
averaged reduced mass $\mu\equiv(\mu_x^2 \mu_z)^{1/3}$ and dielectric constant
$\epss\equiv(\epsilon_{\perp}^2 \epsilon_\parallel)^{1/3}$,
where $1/\mu_i\equiv1/m^e_i+1/m^h_i$ and we use the low-frequency limit of the
dielectric constant tensor of \TDO{} \cite{madelung2000non}:
$\epsilon_{\perp}=\epsilon_x=\epsilon_y=111$
and $\epsilon_\parallel=\epsilon_z=257$.

%%%%%%%%%%%%%%%%%%%%%%%%%
%%%%%%%%%%%%%%%%%%%%%%%%%
\subsection{Dipole and quadrupole matrix elements}
\label{sec_OMAT}
\begin{figure}
	\includegraphics[width=86mm]{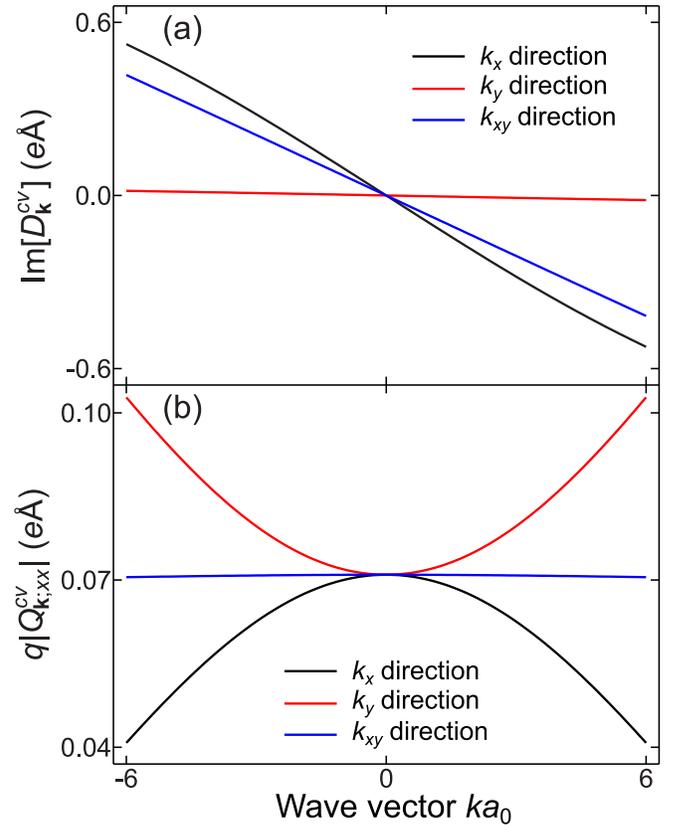}
	\caption{\label{fig2}
	(Color online) Electric-dipole and quadrupole matrix elements. (a) For 
	$x$-polarized light the dipole matrix element $D^{cv}_{\kv}$ is almost linear
	in $k_x$ (black line) and even more linear along the $k_{xy}$ direction
	(blue line) while $D^{cv}_{\kv}$ is negligible along the $k_y$ axis
	(red line). (b) The absolute value of the quadrupole matrix element
	$Q^{cv}_{\kv;xx}$ is constant along the $k_{xy}$ direction (blue line) and
	has parabolic dependence along the $k_x$ (black line) and $k_y$ (red line)
	axes.
	}
\end{figure}

Figure~\ref{fig2}(a) shows the imaginary part of the dipole matrix element
$D^{cv}_{\kv}$ that we obtain from Eq.~(\ref{DME}) with the DFT defined Bloch
functions. The real part of $D^{cv}_{\kv}$ is negligible. We have selected a
$x$-polarized light mode, $\eqv\parallel \hat{\mathbf{x}}$, and then computed
$D^{cv}_{\kv}$ along the $k_x$ axis (black line), $k_y$ axis (red line),
and direction $k(\hat{\kv}_x+\hat{\kv}_y)$ (blue line). In the $k$-space
region of interest, the dipole matrix element becomes almost linearly
dependent on $k_x$. The black line indicates that this dependency is not
completely linear due to the visible nonlinearity for $|ka_0|>3$. In the
$k_{xy}$ direction (blue line), $D^{cv}_{\kv}$ remains linear even far away
from the $\Gamma$ point.

It follows from the linear-in-$k_x$-like dependency that $D^{cv}_{\kv}$ remains
small along the $k_y$ axis (red line) and includes only a minor dependency on
$k_z$ near the $\Gamma$ point. The dipole matrix element for the $y$-polarized
light mode, $\eqv\parallel \hat{\mathbf{y}}$, is essentially equal to the
$\eqv\parallel \hat{\mathbf{x}}$ matrix element with a sign change and with
respect to the $k_x \leftrightarrow k_y$ transformation. In the immediate
vicinity of the $\Gamma$ point, our results indicate that the dipole matrix
element for $z$-polarized light modes, $\eqv\parallel \hat{\mathbf{z}}$, is
negligible, yielding a vanishing near-bandgap absorption in \TDO{} for this
particular light polarization, as already found in several works
\cite{pascual1978fine,amtout1995optical}.
%%%%%%%%%%%%%%%%%%%%%%%%%%%%%%%%%%%%%%%%%%%%%%%%%%
%%%%%%%%%%%%%%%%%%%%%%%%%%%%%%%%%%%%%%%%%%%%%%%%%%

In Fig.~\ref{fig2}(b), we show the absolute value of the $Q^{cv}_{\kv;xx}$ 
element of the quadrupole-matrix tensor multiplied by the wave number
$q=E_g n_o/(\hbar c)$ where $c$ is the speed of light in vacuum and $n_o=2.95$ 
is the ordinary refractive index of \TDO{} \cite{devore1951refractive},
corresponding to a light mode polarized in the $xy$ plane. We have used the
same color scheme as in Fig.~\ref{fig2}(a). In the region of interest,
the $q|Q^{cv}_{\kv,xx}|$ remains constant along the $k_{xy}$ direction
while it has a parabolic behavior along the $k_x$ and $k_y$ axes.

In the vicinity of the $\Gamma$ point, all other components of the quadrupole
and magnetic-dipole matrix-element tensors are at least two orders of magnitude
smaller than $Q^{cv}_{\kv;xx}$. The only exception is the $Q^{cv}_{\kv;yy}$
counterpart of $Q^{cv}_{\kv;xx}$ (obtained from $Q^{cv}_{\kv;xx}$ via the
$k_x \leftrightarrow k_y$ transformation). Due to their small magnitude,
the other elements play only a minor role for the near-bandgap optical
properties of \TDO{}. Nevertheless, we still include them in our numerical
computations. The large difference in the magnitudes of the different
components of the $Q^{ij}_{\kv}$ tensor indicates that the near-bandgap
quadrupole light--matter interaction in \TDO{} should become highly dependent
on the propagation and polarization directions of light. Due to the complicated
nature of the electric-dipole, electric-quadrupole, and magnetic-dipole matrix
elements, we take them as a direct input from the DFT without any
approximations.

%%%%%%%%%%%%%%%%%%%%%%%%%
%%%%%%%%%%%%%%%%%%%%%%%%%
\subsection{Optical absorption}  
\begin{figure}
 \includegraphics[width=86mm]{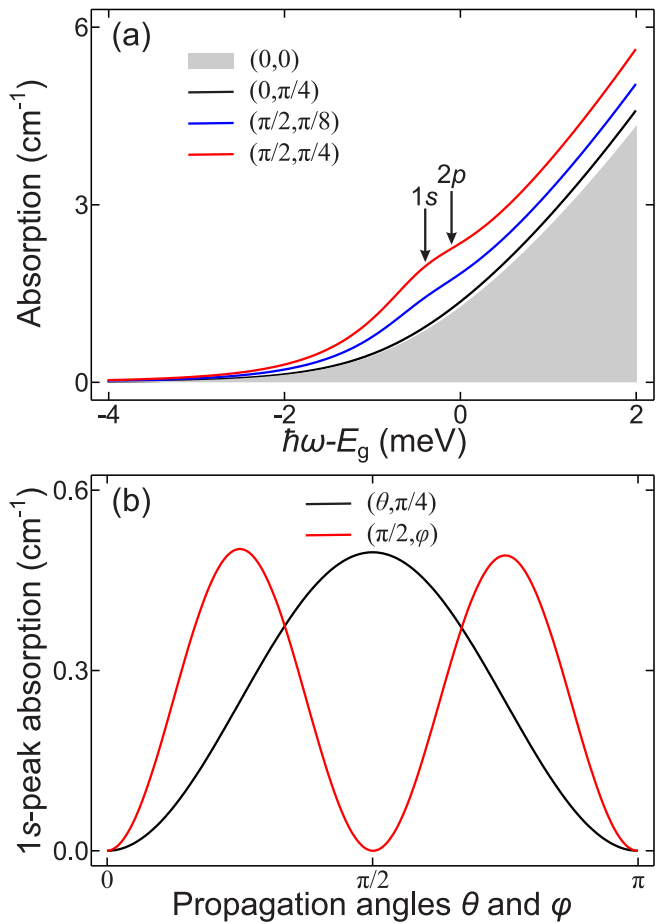}
 \caption{\label{fig3}
 (Color online) Propagation and polarization angle dependency of absorption for
 polarization in $xy$ plane. (a) When light propagates parallel to the z axis
 (shaded area and black line), absorption follows from the electric-dipole
 interaction that does not have strong dependency on the azimuthal angle
 $\varphi$. Changing the propagation direction into the $xy$ plane
 (blue and red curves), the strength of the quadrupole interaction can be tuned
 by varying $\varphi$, yielding notable changes in the spectrum. (b) The
 resonance strength of the $1s$ exciton as a function of the polar angle
 $\theta$ when $\varphi=\pi/4$ (black line) and as a function of $\varphi$
 with a fixed $\theta=\pi/2$ (red line).}
 \end{figure}

Figure~\ref{fig3}(a) shows the optical absorption spectrum following from 
Eq.~(\ref{abso}) when we use the low-frequency values $\epss_\perp=111$ and
$\epss_\parallel=257$ to describe the dielectric screening of Coulomb 
interaction between electrons and holes in Eq.~(\ref{CME}). We have chosen
the polarization of light to be always in the $xy$ plane whereas we change 
the polar $\theta$ and azimuthal $\varphi$ angles that are the angles of
propagation with respect to the $z$ and $x$ axes of the unit cell,
respectively. Here, we present the absorption spectrum for four sets
of $(\theta,\varphi)$ combinations.

Whenever light propagates in the $x$, $y$, or $z$ planes, e.g., in the cases
denoted by $(\theta=0,\varphi=0)$ (shaded area) and $(0,\pi/4)$ (black line)
in Fig.~\ref{fig3}(a), the $Q^{cv}_{\kv;xx}$ and $Q^{cv}_{\kv;yy}$ elements of
the quadrupole matrix-element tensor do not contribute to the light--matter
interaction. Consequently, the absorption spectrum then practically follows
only from the electric-dipole interaction. Since the $D^{cv}_{\kv}$ matrix
elements for $x$- and $y$-polarized light are almost linear in $k_x$ and
$k_y$, respectively, the electric-dipole interaction couples light
predominantly to the $p_{xy}$-like excitonic states of the $xy$ plane
\cite{pascual1977resolved, pascual1978fine, amtout1995optical}.
From the $xy$-plane symmetry of our system, it follows that these $p_{xy}$
states are degenerate and by changing the angles $\theta$ and $\varphi$ we
cannot drastically change their combined contribution to the optical response,
which is shown by the comparison of the spectra labeled $(0,0)$ and
$(0,\pi/4)$ in Fig.~\ref{fig3}(a). Thus, the major propagation- and
polarization-angle dependent changes in our absorption spectrum
(for $xy$ polarization) must follow from the quadrupole interaction.
%%%%%%%%%%%%%%%%%%%%%%%%%%%%%%%%%%%%%%%%%%%%%%%%%%
%%%%%%%%%%%%%%%%%%%%%%%%%%%%%%%%%%%%%%%%%%%%%%%%%%

In Fig.~\ref{fig3}(a), we also show two spectra denoted as $(\pi/2,\pi/8)$
(blue line) and $(\pi/2,\pi/4)$ (red line) in which the propagation direction
points away from the $x$, $y$, or $z$ planes. When comparing the spectra
$(0,0)$ and $(\pi/2,\pi/8)$, we see that while the overall absorption
intensity is slightly increased, a weak resonance at the spectral position of
the $1s$ exciton appears. This transition is dipole-forbidden but
quadrupole-allowed
\cite{pascual1977resolved, pascual1978fine, amtout1995optical}. These trends
are further strengthened when the azimuth angle is increased to
$\varphi=\pi/4$, yielding the $(\pi/2,\pi/4)$ set with the strongest
quadrupole resonance.

In Fig.~\ref{fig3}(b), we study how the peak intensity of the $1s$ state
changes as a function of the angles $\theta$ and $\varphi$ that give the
spectra labeled as $(\theta,\pi/4)$ (black line) and $(\pi/2,\varphi)$
(red line), respectively. More generally, whenever $\theta$ ($\varphi$)
is varied we find that $\varphi = \pi/4$ ($\theta = \pi/2$) gives the
strongest $1s$ intensity. As a summary, we show the angle dependency of the
$1s$-peak intensity in Fig.~\ref{fig3}(b), varying only one angle at a time
while fixing the other. The shape of these curves follows almost identically
the corresponding angular dependency of $|Q^{cv}_{00}|^2$.
 
Even with the most optimal angle combination $(\pi/2,\pi/4)$ for the quadrupole
interaction, the resulting excitonic absorption signature is still weaker than
the experimentally observed feature \cite{pascual1977resolved, pascual1978fine,
amtout1992time, amtout1992resonant, amtout1995optical}. This is explained by
the extremely small binding energy of $E^B_{1s}=0.5$~meV
($E^B_{2p_{xy}}=0.1$~meV) for the $1s$ ($2p_{xy}$) exciton due to the unusually
large dielectric constants $\epss_\perp=111$ and $\epss_\parallel=257$.
Additionally, the experimentally measured continuum absorption intensity
around 2~meV above the band gap has been reported to be between 25 cm$^{-1}$
and 50 cm$^{-1}$ \cite{pascual1977resolved, pascual1978fine, amtout1995optical},
i.e., our result differ from this roughly by a factor of five.

Generally, it is not obvious whether we should use a low-frequency, a
high-frequency or some other phenomenological value of a dielectric constant
when we model optical excitations in semiconductors \cite{bechstedt2015many}.
This is of particularly significance in \TDO{} where the difference between
high-frequency constants \cite{madelung2000non} ($\epss_\perp^{\infty}=6.8$ 
and $\epss_\parallel^{\infty}=8.4$) and low-frequency constants
($\epss\perp^0=111$ and $\epss\parallel^0=257$) is exceptionally large.
In rutile, the contributions of the different sources to the
screening of the Coulomb interaction, e.g., phonons and the ionic nature
of \TDO{} \cite{persson2005strong}, are poorly known \cite{chiodo2010self}.

\begin{figure}
\includegraphics[width=86mm]{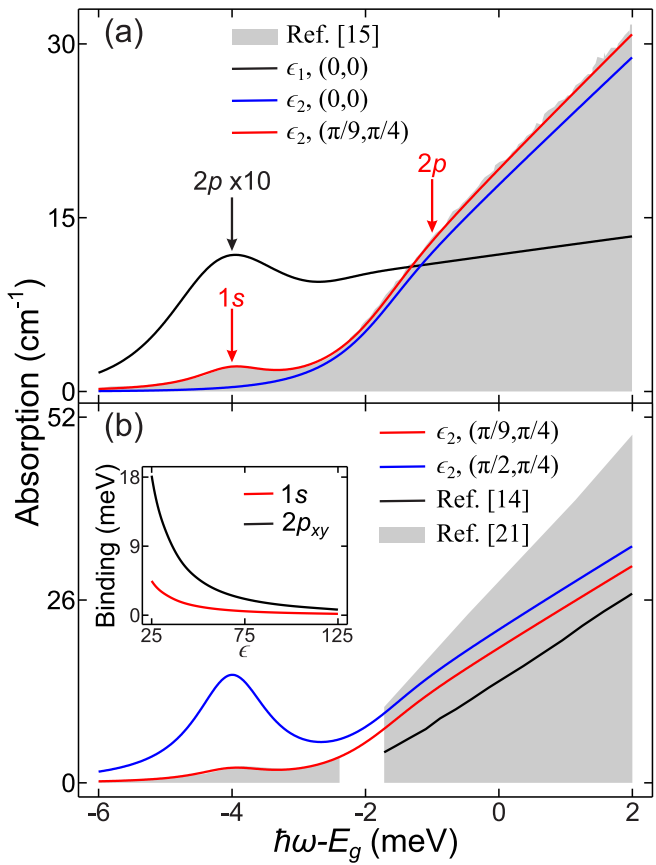}
\caption{\label{fig4}
(Color online) Experiments vs. modeled absorption.
(a) Computed absorption spectrum with $\epss(w)=24.3\equiv \epss_1$
(black line, scaled down by dividing with $10$) has a drastically different
general shape than what is observed in the experiments of 
Ref.~\cite{amtout1995optical} (shaded area) and its intensity is one order of
magnitude too strong. With $\epss(w)=47.8=\epss_2$ and light
propagation parallel to $z$ axis (blue line), the shape and magnitude of
the continuum absorption corresponds to the experiment. Changing
the propagation to $\theta=\pi/9$ and $\varphi=\pi/4$ (red line), we can
achieve similar excitonic feature as in the experiment via $1s$ exciton. 
(b) Spectrum computed with $\epss_2$, $\theta=\pi/9$, and $\varphi=\pi/4$
(red line) also agrees well with other experiments in
Ref.~\cite{pascual1977resolved} (black line) and in Ref.~\cite{pascual1978fine}
(shaded area). With the angle 
that supports the most strongest quadrupole interaction (blue line), we find
a clear $1s$-exciton resonance. (inset) The $1s$ and $2p_{xy}$ binding energies
as a function of $\epss$.
}
\end{figure}

If we adjust the components of the dielectric constant between the high and
low frequency values via $\epss_i(w)=\epss^0_i-(\epss^0_i-\epss^\infty_i)w $,
where $w$ is between 0 and 1, we can change the binding energies of excitons so
that they become visible in the optical spectra. In the inset of
Fig~\ref{fig4}(b), we show how the binding energies of $1s$ (black line)
and $2p$ (red line) change as a function of
$\epss(w)=[\epss_\perp^2(w)\epss_\parallel(w)]^{1/3}$.
By selecting either $\epss(w)=24.3$ or $\epss(w)=47.8$, we can
tune the absorption to show a distinct resonance 4~meV below the band gap.
Effectively, we have adjusted the binding energies of either the $2p_{xy}$ 
or $1s$ state being the origin of that signature.
%%%%%%%%%%%%%%%%%%%%%%%%%%%%%%%%%%%%%%%%%%%%%%%%%%%
%%%%%%%%%%%%%%%%%%%%%%%%%%%%%%%%%%%%%%%%%%%%%%%%%%%

In Fig.~\ref{fig4}, we compare absorption spectra for different
$(\theta,\varphi)$ sets of propagation angles and dielectric constants
to the available experiments.
If we choose $(0,0)$ and $\epss(w)=24.3$ [black line in Fig.~\ref{fig4}(a)],
the experimental resonance from Ref.~\cite{amtout1995optical} (shaded area)
is assigned to the $2p$ exciton. However, the general shape of the spectrum
does not correspond to the measured one, i.e., the $2p$ resonance is too
strong compared to the continuum absorption. This trend is strengthened
when $\epss(w)$ is decreased. In general, whenever the $2p$ resonance is
clearly separable from the continuum tail, its ratio to the continuum
absorption is too high.
 
Our analysis shows that the shape of the full bandedge absorption spectrum
does not match the experimental observations if we try to model the excitonic
signature via the $2p$ state, based on three compelling facts. First, if we
assume that the band energies can be described in the form of
Eq.~(\ref{parabolic}), $\epss(w)$ and the effective masses become connected
via Eq.~(\ref{wannier_rel}). If we change the value of $\epss(w)$, we not only
describe the screening of electron--electron interaction phenomenologically
but also correct possible errors in our effective masses. Second, without
considering $\epss(w)$, we can assume that our Coulomb matrix element is highly
accurate and cannot contribute to this problem, as discussed in
Appendix~\ref{Cint}. Hence, the only remaining uncertainty is in the
light--matter matrix elements. Thirdly, since both the $2p$-resonance and
the continuum absorption (dominantly) originate from the same source
(dipole interaction), possible errors in the dipole matrix element
$D^{cv}_{\kv}$ would lead only to the scaling of both features with the same
factor without a change of the general shape of the spectrum. Thus, based on
our results it seems to be practically impossible to model the experimentally
observed excitonic feature via $2p$ states.

No such problems occur if we model the excitonic resonance as the $1s$ state
by selecting $\epss(w)=47.8$, yielding $E^B_{1s}=4.0$~meV and
$E^B_{2p_{xy}}=1.0$~meV. Interestingly, the experimentally obtained
\cite{pascual1977resolved, pascual1978fine} binding-energy difference between
$1s$ and $2p$ states of 3~meV is well reproduced with these values. Further,
selecting the propagation angles $(0,0)$ [blue line in Fig~\ref{fig4}(a)], we
find that our continuum absorption has the same shape and intensity as in the
experiment with a high accuracy. By changing the propagation angle, we can tune
the intensity of the $1s$ resonance without drastically changing the shape of
the continuum absorption. With a rather small angle of $\theta=\pi/9$ and the
optimal $\varphi=\pi/4$ (red line), we find an extremely good agreement to the
spectrum of Ref.~\cite{amtout1995optical}.

We also compare our results against other experiments from
Refs.~\cite{pascual1977resolved} and \cite{pascual1978fine} in
Fig.~\ref{fig4}(b).
Again, we show $(\pi/9,\pi/4)$ with $\epss(w)=47.8$ (red line) and additionally
$(\pi/2,\pi/4)$ with $\epss(w)=47.8$ (blue line).
Once more, the excitonic resonance is reproduced with great precision.
The continuum absorption of both sets settles in-between the ones from
Ref.~\cite{pascual1977resolved} (black line) and Ref.~\cite{pascual1978fine}
(shaded area). As previously, the quadrupole interaction of the $(\pi/2,\pi/4)$
produces the highest intensity for the $1s$ resonance that is now roughly seven
times stronger than in the experiments.

In Fig.~\ref{fig4}, the band gaps for experimental data have been fixed so
that the spectral position of the excitonic resonance matches with the result
of Ref.~\cite{amtout1995optical}, where the resonance is located at
$\hbar\omega=3.03$~eV. For our theoretical results, we do essentially the same
by using the scissor shifted $E_g=3.034$~eV. In results of Figs.~\ref{fig3}
and \ref{fig4}, we have selected the scattering function
$\gamma_{\nu}(\omega)$ in Eq.~(\ref{dephasing}) to be
$\gamma_{1s}(\omega)=0.8$~meV for our $1s$ states, producing a similar shape
and linewidth broadening for the modeled $1s$ resonance as seen for the
experimental resonance. For all the other excitonic states, we use
$\gamma_\nu=1.8$~meV and $\mu_\nu=\Delta E_\nu=1.2$~meV. With these selections
we mimic the excitation-induced dephasing \cite{PQE06, smith2010extraction},
without drastically altering location and intensities of our resonances
compared to a case $\gamma_{\nu}(\omega)=\gamma_{\nu}$, but reproduce the
experimentally detected \cite{tang1995urbach} steep decay of the continuum
absorption tail in rutile \TDO{}.

%%%%%%%%%%%%%%%%%%%%%%%%%
%%%%%%%%%%%%%%%%%%%%%%%%%
\section{Summary}
In summary, we systematically combine DFT and the cluster-expansion approach
to compute the excitonic optical properties of semiconductors. More
specifically, the matrix elements needed for microscopic modeling are
computed via DFT while the many-body problem is solved with the cluster
expansion. We apply this hybrid approach for rutile \TDO{} and compute
its near-bandgap optical absorption, following from the electric-dipole,
electric-quadrupole and magnetic-dipole light--matter interactions.
Our results show that the quadrupole interaction in rutile is highly dependent
on the propagation and polarization direction of light. Furthermore, we find
that it is hard to explain the experimentally detected excitonic signature in
the absorption spectrum of \TDO{} by considering only electric-dipole
interaction and the dipole-allowed $2p$ exciton. We obtain an excellent
agreement between modeled and experimental spectra through the quadrupole
interaction and the quadrupole-allowed $1s$ exciton if the light is propagating
in a sufficiently large angle with respect to the crystallographic axes of
\TDO{}. Hence, the hybrid CE-DFT method seems to be a very promising approach
to model many-body effects in semiconductors, opening a wide range of new
possibilities to study and utilize properties of nontrivial systems.

\section{Acknowledgment}
We thank John E. Sipe for helpful discussions. This work is funded by the
Deutsche Forschungsgemeinschaft via the SFB 1083. DSP also acknowledges
support from  Spanish MINECO (Grant MAT2013-46593-C6-2-P).

%%%%%%%%%%%%%%%%%%%%%%%%%%%%%%%%%%%%%%%%%%%%%%%%%%
%%%%%%%%%%%%%%%%%%%%%%%%%%%%%%%%%%%%%%%%%%%%%%%%%%
%%%%%%%%%%%%%%%%%%%%%%%%%%%%%%%%%%%%%%%%%%%%%%%%%%
%%%%%%%%%%%%%%%%%%%%%%%%%%%%%%%%%%%%%%%%%%%%%%%%%%
%%%%%%%%%%%%%%%%%%%%%%%%%%%%%%%%%%%%%%%%%%%%%%%%%%
%%%%%%%%%%%%%%%%%%%%%%%%%%%%%%%%%%%%%%%%%%%%%%%%%%
\appendix
\section{Renormalized terms}
\label{RQ}
The exact form of the renormalized kinetic energy $\rene^{\la}_{ij}$ and
the renormalized Rabi frequency $\Omega^{\la}_{ij}$ appearing in 
Eq.~(\ref{SBE}) are given by
\begin{align}
\label{rene}
	\rene^{\la}_{ij} = & \delta_{ij}\spe{\la}{i} -E(t)D^{\la\la}_{ji}
	+ \sum_{k,l} \Big[{\mathcal{V}}^{\la\la;\la\la}_{jk;li} P^{\la\la}_{kl}
	+{\mathcal{V}}^{\la\bar{\la};\bar{\la}\la}_{jk;li}  
	P^{\bar{\la}\bar{\la}}_{kl} 
	\nonumber \\
	& +{\mathcal{V}}^{\la\la;\bar{\la}\la}_{jk;li}  P^{\la\bar{\la}}_{kl}
	+{\mathcal{V}}^{\la\bar{\la};\la\la}_{jk;li}  P^{\bar{\la}\la}_{kl} \Big]
	-{\mathcal{V}}^{E;\la}_{ji} \, , \\
%%%%%
%%%%%
\label{rabi}
	\Omega^{\la}_{ij} = & E(t)D^{\bar{\la}\la}_{ji}- \sum_{k,l} 
	\Big[{\mathcal{V}}^{\bar{\la}\la;\bar{\la}\la}_{jk;li} P^{\la\bar{\la}}_{kl}
	+{\mathcal{V}}^{\bar{\la}\la;\la\la}_{jk;li} P^{\la\la}_{kl} \nonumber \\
	&+{\mathcal{V}}^{\bar{\la}\bar{\la};\bar{\la}\la}_{jk;li} 
	P^{\bar{\la}\bar{\la}}_{kl}
	+{\mathcal{V}}^{\bar{\la}\bar{\la};\la\la}_{jk;li}
	P^{\bar{\la}\la}_{kl}\Big] + {\mathcal{V}}^{\Omega;\la}_{ji},
\end{align}
where
\begin{align*}
	{\mathcal{V}}^{E;v}_{ji} & \equiv \sum_k {\mathcal{V}}^{vv;vv}_{jk;ki},
	\qquad {\mathcal{V}}^{E;c}_{ji} \equiv \sum_k {\mathcal{V}}^{cv;vc}_{jk;ki},
	\nonumber \\
	{\mathcal{V}}^{\Omega;v}_{ji} & \equiv \sum_k {\mathcal{V}}^{cv;vv}_{jk;ki},
	\qquad
	{\mathcal{V}}^{\Omega;c}_{ji} \equiv \sum_k {\mathcal{V}}^{vv;vc}_{jk;ki},
\end{align*}
have been added to ensure that the kinetic energy~(\ref{rene}) and Rabi
frequency~(\ref{rabi}) of the ground state will not become renormalized.
By adding the corrections ${\mathcal{V}}^{E;\la}_{ji}$ and
${\mathcal{V}}^{\Omega;\la}_{ji}$ into Eqs.~(\ref{rene}) and (\ref{rabi}),
we avoid double counting contributions that are already included in the
effective single-particle potential $U(\rv)$ in Eq.~(\ref{spH}).

As we evaluate Eq.~(\ref{SBE}) for the polarization $P_{ij}$, we find that
all the electron--electron-interaction terms of the form
${\mathcal{V}}^{\la\la;\la\bar{\la}}_{jk;li}$ in Eqs.~(\ref{rene}) and
(\ref{rabi}) as well as $E(t)D^{\la\la}_{ji}$ in Eq.~(\ref{rene}) are
related to a nonlinear response. Consequently, these terms do not contribute
to the linear absorption. Neglecting the
${\mathcal{V}}^{\bar{\la}\bar{\la};\la\la}_{jk;li}$ term in Eq.~(\ref{rabi})
that couples $P^{vc}_{ij}$ and $P^{cv}_{ij}$ polarizations is
known as the Tamm-Dancoff approximation \cite{Onida_Reining:2002}.
By setting $f^e_{ij}=f^h_{ij}=0$, we obtain the final form of Eq.~(\ref{PSBE}).

These neglected ${\mathcal{V}}$ terms originate from processes where
electron--electron interaction scatters electronic states between valence- and
conduction-band states. When they are separated by a sufficiently large energy
gap, this scattering is energetically highly unfavorable. Hence, all terms of
the form $V^{\la\la;\la\bar{\la}}_{jk;li}$,
$V^{\la\la;\bar{\la}\bar{\la}}_{jk;li}$, and
$V^{\la\bar{\la};\la\bar{\la}}_{jk;li}$ in Eqs.~(\ref{rene}) and
(\ref{rabi}) can typically be omitted, which drastically simplifies the
problem and reduces numerical efforts.

For crystalline solids (with sufficient $E_g$), these terms remain negligible
as long as the system is excited in a sufficiently small region of the
Brillouin zone (see Appendix~\ref{Cint}). However, the effect of these terms
should be verified based on the studied properties of any new system, as we do
in Appendix~\ref{Cint} for the near-bandgap optical properties of rutile
\TDO{}. For example, it is found that the Tamm-Dancoff approximation can fail
in molecular systems \cite{doi:10.1021/nl803717g} while in some cases it can
improve the results \cite{doi:10.1021/ct200651r}.

%%%%%%%%%%%%%%%%%%%%%%%%%
%%%%%%%%%%%%%%%%%%%%%%%%%
\section{Light--matter matrix elements}
\label{LM}
In an infinite crystal, the $r_i$-matrix element
of the $i$-th spatial component of $\rv$ becomes
ambiguously defined via normal functions, but can be expressed via 
distributions like the Dirac delta function and its derivatives 
\cite{blount1962formalisms, gu2013relation}.
As we show later, the same holds for products of $r_i$ operators.
In the scope of this work, we can properly understand these generalized
functions by considering the oscillator strength, Eq.~(\ref{oss}), that is
given in a two-band model by an integration over the first Brillouin zone
\begin{equation}
\label{ossCS}
	{\mathcal{F}}_\nu = \int_{BZ} d^3 k^c d^3 k^v \phi^*_{\nu} (\kv^v,\kv^c)
	F^{cv}_{\kv^c\kv^v},
\end{equation}
and the connection between bands $\la=c$ and $\la=v$ to the wave vectors
$\kv^c$ and $\kv^v$ is explicitly denoted. When the excitation of the system
remains inside a sufficiently small region of the Brillouin zone, the
integration boundaries can be extended to infinity since the excitonic wave
functions approach zero. Assuming a parabolic energy
dispersion~(\ref{parabolic}), we can introduce a change of variables in 
Eq.~(\ref{ossCS}) to the center-of-mass, $\qv$, and
relative-movement, $\kv$, wave vectors that are related to the wave vectors
$\kv^c$ and $\kv^v$ by the transformation
\begin{equation}
	k^c_i=k_i+\frac{m^e_i}{M_i}q_i, \quad k^v_i=k_i-\frac{m^h_i}{M_i}q_i.
\end{equation}
Equation~(\ref{ossCS}) then reads
\begin{equation}
\label{ossCS_COM}
	{\mathcal{F}}_\nu = \int d^3 q \, d^3 k \, \phi^*_{\nu,\qv} (\kv)
	F^{cv}_{\kv^c\kv^v}.
\end{equation}

We can give all terms in Eq.~(\ref{MPoper}) through operators
$\prod_{i=1}^3 r_i^{n_i}$ and $\big(\prod_{i=1}^3 r_i^{n_i}\big)p_j$
where $n_i$ is a non-negative integer. The matrix elements for these
operators are given by
\begin{align}
\label{RME}
	{\mathcal{R}}^{\mathbf{n}}_{\kv^c\kv^v} \equiv & \int d^3 r\, 
	\psi^*_{c,\kv^c}(\rv) \Bigg(\prod_{i=1}^3 r_i^{n_i} \Bigg) 
	\psi_{v,\kv^v}(\rv) 
	\nonumber \\
	=&\sum_\Gv U^{cv}_{\kv^c\kv^v}(\Gv)\frac{1}{(2\pi)^3}\int d^3 r \,
	e^{-i (\qv-\Gv)\cdot\rv}  \prod_{i=1}^3 r_i^{n_i} 
	\nonumber \\
	\equiv&\langle u_{c,\kv^c} | u_{v,\kv^v} \rangle_{\Omega_0} \Bigg(
	\prod_{i=1}^3 i^{n_i}\frac{\partial^{n_i}}
	{\partial q_i^{n_i}}\Bigg)\delta(\qv),
	\\
\label{PME}
	{\mathcal{P}}^{\mathbf{n};j}_{\kv^c\kv^v} \equiv & \int d^3 r\, 
	\psi^*_{c,\kv^c}(\rv) \Bigg(\prod_{i=1}^3 r_i^{n_i} \Bigg) p_j 
	\psi_{v,\kv^v}(\rv)
	\nonumber \\
	\equiv & \langle u_{c,\kv^c} | p_j | u_{v,\kv^v} \rangle_{\Omega_0} 
	\Bigg(\prod_{i=1}^3 i^{n_i}\frac{\partial^{n_i}}{\partial q_i^{n_i}}\Bigg)
	\delta(\qv) 
	\nonumber \\ 
	&+ \hbar\Bigg(k_j-\frac{m^h_j}{M_j}q_j\Bigg)
	{\mathcal{R}}^{\mathbf{n}}_{\kv^c\kv^v},
\end{align}
where $\mathbf{n}=(n_1,n_2,n_3)$ defines the integers $n_i$, and we have assumed
that the excitation remains inside a region where $|\qv|<|\Gv|$ for any nonzero
reciprocal-lattice vector $\Gv$.
We also use the unit-cell Fourier transforms for
$u^*_{c,\kv^c}(\rv) u_{v,\kv^v}(\rv)$ and 
$u^*_{c,\kv^c}(\rv) p_j u_{v,\kv^v}(\rv)$
[see Eqs.~(\ref{UFT1}) and (\ref{UFT2})] and
\begin{equation}
	\langle  u_{\la,\kv} |  \hat{\mathcal{O}} |u_{\la',\kv'} \rangle =
	\frac{1}{\Omega_0} \int_{\Omega_0} d^3 r\, u^*_{\la,\kv}(\rv) 
	\hat{\mathcal{O}} u_{\la',\kv'}(\rv)
\end{equation}
denotes the matrix element of an operator $\hat{\mathcal{O}}$.

The oscillator strength (\ref{ossCS_COM}) is related to the matrix elements
in Eqs.~(\ref{RME}) and (\ref{PME}) via
\begin{align}
\label{ossR_o}
	{\mathcal{F}}^{\mathbf{n}}_\nu \equiv &\int d^3 q \, d^3 k \,\phi^*_{\nu,\qv}
	(\kv) {\mathcal{R}}^{\mathbf{n}}_{\kv^c\kv^v} = \int d^3 k \, \phi^*_{\nu,0}
	(\kv)  {\mathcal{R}}^{\mathbf{n}}_{\kv},
	\\
\label{ossP_o}
	{\mathcal{F}}^{\mathbf{n};j}_\nu \equiv &\int d^3 q \, d^3 k \, 
	\phi^*_{\nu,\qv} (\kv) {\mathcal{P}}^{\mathbf{n};j}_{\kv^c\kv^v} 
	\nonumber \\
	= & \int d^3 k \, \phi^*_{\nu,0} (\kv)  ({\mathcal{P}}^{\mathbf{n};j}_{\kv}
	+\hbar k_j {\mathcal{R}}^{\mathbf{n}}_{\kv}-\hbar\frac{m^h_j}{M_j}
	\tilde{{\mathcal{R}}}^{\mathbf{n};j}_{\kv}),
\end{align}
where
\begin{align}
\label{ossR}
	{\mathcal{R}}^{\mathbf{n}}_{\kv} \equiv & \Big(\Pi_{i=1}^3 (-i)^{n_i}
	\frac{\partial^{n_i}}{\partial q_i^{n_i}}\Big) 
	\langle u_{c,\kv^c} | u_{v,\kv^v} \rangle_{\Omega_0}\Big|_{\qv=0},
	\\
\label{ossP}
	{\mathcal{P}}^{\mathbf{n};j}_{\kv} \equiv & \Big(\Pi_{i=1}^3 (-i)^{n_i}
	\frac{\partial^{n_i}}{\partial q_i^{n_i}}\Big)
	\langle u_{c,\kv^c} |p_j| u_{v,\kv^v} \rangle_{\Omega_0}\Big|_{\qv=0},\\
\label{ossRt}
	\tilde{{\mathcal{R}}}^{\mathbf{n};j}_{\kv} \equiv & \Big(\Pi_{i=1}^3
	(-i)^{n_i}\frac{\partial^{n_i}}{\partial q_i^{n_i}}\Big) q_j 
	\langle u_{c,\kv^c} | u_{v,\kv^v} \rangle_{\Omega_0}\Big|_{\qv=0}.
\end{align}
In general, a Bloch function $u_{\la,\kv}(\rv)$ is an analytic function of
$\kv$ if the band $\la$ is not degenerate
\cite{PhysRevLett.54.1075, 0953-8984-12-34-201},
yielding a well-defined differentiation in Eqs.~(\ref{ossR})-(\ref{ossRt}).

A comparison between Eqs.~(\ref{ossR_o}), (\ref{ossP_o}), and (\ref{ossCS})
reveals that the matrix element in Eqs.~(\ref{RME}) and (\ref{PME}) are
effectively direct in $k$ space, i.e., their contribution to the
light--matter interaction are given by
\begin{align}
\label{RME_D}
	{\mathcal{R}}^{\mathbf{n}}_{\kv^c\kv^v} = & 
	\delta(\kv^c-\kv^v)  {\mathcal{R}}^{\mathbf{n}}_{\kv^c} \, , \\
\label{PME_D}
	{\mathcal{P}}^{\mathbf{n};j}_{\kv^c\kv^v} = &
	\delta(\kv^c-\kv^v) {\mathcal{P}}^{\mathbf{n};j}_{\kv^c} 
	+\hbar k_j {\mathcal{R}}^{\mathbf{n}}_{\kv^c}
	-\hbar\frac{m^h_j}{M_j}\tilde{{\mathcal{R}}}^{\mathbf{n};j}_{\kv^c},
\end{align}
and can be determined after Eqs.~(\ref{ossR})-(\ref{ossRt}) have been solved.
For nondegenerate bands $c$ and $v$, we can do this systematically by
applying $\kv\cdot\pv$ theory and the conventional $n^\text{th}$-order
nondegenerate perturbation theory by expanding the functions $u_{c,\kv^c}$
and $u_{v,\kv^v}$ at $\kv$ where $n$ is the largest $n_i$ within
Eqs.~(\ref{ossR})-(\ref{ossRt}).

In this work, we consider the electric-dipole, electric-quadrupole,
and magnetic-dipole interactions in Eqs.~(\ref{Dope})-(\ref{Mope}) that are
described by
\begin{align}
\label{DopeRP}
	D = &-e \sum_i e_i r_i,  \\ 
\label{QopeRP}
	Q = & -i \frac{e}{2} \sum_{i,j} e_i q_j r_i r_j ,  \\
\label{MopeRP}
	M = & -\frac{e}{2\omega m_0}\sum_{i,j}(e_i q_j-e_j q_i) r_i p_j.
\end{align}
We obtain the matrix elements in Eqs.~(\ref{DME})-(\ref{MME}) by using
Eqs.~(\ref{ossR})-(\ref{PME_D}), $\kv\cdot\pv$ theory, and the second order
perturbation theory for matrix elements of operators in
Eqs.~(\ref{DopeRP})-(\ref{MopeRP}).
This rater lengthy but straightforward calculation yields 
${{Q}}^{cv}_{\kv;ij}=\tilde{{{Q}}}^{cv}_{\kv;ij}
+\tilde{{{Q}}}^{cv}_{\kv;ji}$ and ${{M}}^{cv}_{\kv;ij}=
\tilde{{{M}}}^{cv}_{\kv;ij}-\tilde{{{M}}}^{cv}_{\kv;ji}$
with
\begin{align}
	\tilde{{{Q}}}^{cv}_{\kv;ij} = & -i \frac{e}{2} \Big(1
	-\frac{m^e_j m^h_i}{M_i M_j} \Big) q^{cv}_{\kv;ij}
	\nonumber \\
	& + i\frac{e\hbar}{2m_0}  [M_i M_j(\spe{c}{\kv}-\spe{v}{\kv})]^{-1}
	\Big[ m^h_i m^h_j[ s^{cv}_{\kv;ij} 
	\nonumber \\
	&- p^{cc}_{\kv;i}\tilde{p}^{cv}_{\kv;j}] -m^e_i m^e_j[ s^{vc}_{\kv;ij}
	-p^{vv}_{\kv;i}\tilde{p}^{vc}_{\kv;j}]^* \Big], \\
%%%%%%%%%%%%%%%%%%%%%
	\tilde{{{M}}}^{cv}_{\kv;ij} = & -i \frac{e}{2\omega m_0} \Big[ \hbar k_j
	\tilde{p}^{cv}_{\kv;i} + \frac{m^e_i}{M_i} s^{cv}_{\kv;ij} 
	-\frac{m^h_i}{M_i} (s^{vc}_{\kv;ij})^* \Big].
\end{align}
Here, we have defined
\begin{align}
q^{\la\la'}_{\kv;ij} \equiv & \sum_{\eta\neq \la,\la'} 
\tilde{p}^{\la\eta}_{\kv;i} \tilde{p}^{\eta\la'}_{\kv;j}, %\\
\quad s^{\la\la'}_{\kv;ij} \equiv & \sum_{\eta\neq \la} 
\tilde{p}^{\la\eta}_{\kv;i} p^{\eta\la'}_{\kv;j} \, ,
\end{align}
where $p^{\la\la'}_{\kv;i}$ ($\tilde{p}^{\la\la'}_{\kv;i}$) is the 
$i^\text{th}$ component of the vector
 $\pv^{\la\la'}_{\kv}$ ($\tilde{\pv}^{\la\la'}_{\kv}$).

%%%%%%%%%%%%%%%%%%%%%%%%%
%%%%%%%%%%%%%%%%%%%%%%%%%
\section{Coulomb interaction in crystals}
\label{Cint}
In a crystal with electronic wave functions~(\ref{SPWF}) and electron--electron
interaction~(\ref{CI})-(\ref{CME}), the explicit Coulomb matrix element
in Eq.~(\ref{GCME}) becomes
\begin{align}
\label{CME_exa}
	V^{\la_1 \la_2; \la_3 \la_4}_{\kv_1\kv_2;\kv_3\kv_4}= & \sum_{\Gv, \Gv'}
	U^{\la_1\la_4}_{\kv_1\kv_4}(\Gv) U^{\la_2\la_3}_{\kv_2\kv_3}(\Gv')
	V_{\kv_3-\kv_2+\Gv'}
	\nonumber \\
	&\times \delta(\Gv+\Gv'-\kv_1-\kv_2+\kv_4+\kv_3) \, ,
\end{align}
where we have expressed the Fourier transform of pairs of Bloch functions
as a sum
\begin{align}
\label{UFT1}
	&u^*_{\la,\kv}(\rv) u_{\la',\kv'}(\rv)  = \sum_{\Gv} 
	U^{\la\la'}_{\kv\kv'}(\Gv) e^{i\Gv\cdot\rv},
	\\
\label{UFT2}
	&U^{\la\la'}_{\kv\kv'}(\Gv) = \frac{1}{\UC} \int_{\UC}
	d^3r \, u^*_{\la,\kv}(\rv) u_{\la',\kv'}(\rv) e^{-i\Gv\cdot\rv},
\end{align}
over the reciprocal lattice vectors $\Gv$.
If we now consider a system where the excitation remains inside a region where
$|\kv_1+\kv_2-\kv_4-\kv_3|<|\Gv|$ for any nonzero $\Gv$, Eq.~(\ref{CME_exa})
produces
\begin{equation}
\label{CCME}
	V^{\la_1 \la_2; \la_3 \la_4}_{\kv_1\kv_2;\kv_3\kv_4} = 
	\delta(\kv_1+\kv_2-\kv_3-\kv_4) 
	V^{\la_1 \la_2; \la_3 \la_4}_{\kv_4\kv_3;\kv_1-\kv_4},
	\end{equation}
where
\begin{align}
\label{CME_comp}
	V^{\la_1 \la_2; \la_3 \la_4}_{\kv\kv';\qv} =  
	&\tilde{V}^{\la_1 \la_2; \la_3 \la_4}_{\kv \kv';\qv}
	+ \sum_{\Gv\neq 0}  U^{\la_1\la_4}_{(\kv+\qv)\kv}(\Gv) \nonumber \\
	& \times U^{\la_2\la_3}_{(\kv'-\qv)\kv'}(-\Gv) V_{\qv-\Gv}
\end{align}
contains the matrix element
\begin{align}
\label{CME_nonUK}
	& \tilde{V}^{\la_1 \la_2; \la_3 \la_4}_{\kv\kv';\qv}
	=  V_{\qv}\langle u_{\la_1,\kv+\qv} | u_{\la_4,\kv}
	\rangle_{\Omega_0}\langle u_{\la_2,\kv'-\qv} | u_{\la_3,\kv'}
	\rangle_{\Omega_0}.
\end{align}

Since $\tilde{V}^{\la_1 \la_2; \la_3 \la_4}_{\kv\kv';\qv}$ diverges
at $\qv=0$, the leading Coulomb-interaction contributions will follow
from this term. The actual $\qv=0$ divergence can be removed by using the
jellium model \cite{haug2009quantum}.We next consider excitations where
$|\qv| \ll |\Gv|$ for any nonzero $\Gv$ so that we can approximate
$\langle u_{\la,\kv} | u_{\la',\kv+\qv} \rangle_{\Omega_0}
= \delta_{\la\la'}$. If we only allow contributions that are proportional
to $1/|\qv|$, the matrix element~(\ref{SC_CME}) directly follows
from~(\ref{CME_comp}) (see Ref.~\cite{bechstedt2015many} for a similar
approach).

We next check the approximated Coulomb matrix elements for \TDO{}.
To do this, we consider the most important direct elements given by
$V^{v c; c v}_{\kv\kv;\qv}$. For these terms we study how well the
approximate Coulomb matrix element $V_\qv$ compares with the explicit 
$V^{v c; c v}_{\kv\kv;\qv}$ obtained from DFT.
By performing a great number of control computations using $k$ grids
with $|\kv|a_0<10$ and $|\qv|a_0<10$, we find a maximal error of
only 3\% while in the most important region for the convergence of
excitonic states the error is less than $10^{-3}$.

We also have computed the matrix elements
${\mathcal{V}}^{vv;cc}_{\kv\kv;(\kv-\qv)(\kv+\qv)}$ and
$V^{vc;vc}_{\kv\kv;(\kv-\qv)(\kv+\qv)}$ that contribute to the linear
response if the Tamm-Dancoff approximation is not made and the exchange terms
in Eqs.~(\ref{PSBE})-(\ref{CandE}) are not omitted. In the region of interest,
these Coulomb contributions are negligible compared to the leading Coulomb
terms and orders of magnitude smaller than $E_g$. Hence, it is justified to
omit these terms.

%%%%%%%%%%%%%%%%%%%%%%%%%
%%%%%%%%%%%%%%%%%%%%%%%%%
\section{Excitonic states}
\label{Xs}

Excitonic states are defined by the Wannier equation~(\ref{wannier}).
Using the effective mass approximation Eq.~(\ref{parabolic}), the
Coulomb matrix elements Eq.~(\ref{CME}), and the two dielectric
constants of \TDO{}, it is given by
\begin{align}
\label{IWE}
\Big\{ & \frac{\hbar^2}{2 \mu_{\perp}}\kv_\perp^2+\frac{\hbar^2}
{2 \mu_\parallel} k_z^2 \Big\} \phi_{\la}(\kv)
- \frac{e^2}{4\pi^3\varepsilon_0\epsilon_\perp} \int d^3k' \phi_{\la}(\kv')
\nonumber \\
&\times \Big[(\kv_\perp-\kv'_\perp)^2+\frac{\epsilon_\parallel}
{\epsilon_\perp}(k_z-k_z')^2 \Big]^{-1} = E_{\la}\phi_{\la}(\kv) \, ,
\end{align}
where the momentum $\kv \equiv (\kv_\perp,k_z)$ has been decomposed into
directions in ($\kv_\perp$) and out-of ($k_z$) the $k_{xy}$ plane.
It is beneficial to make the coordinate transformation 
$ k_z \rightarrow \sqrt{\epsilon_\perp/\epsilon_\parallel}k_z $
that converts Eq.~(\ref{IWE}) into
\begin{align}
\label{IWET}
\Big\{ & \frac{\hbar^2}{2 \mu_{\perp}}\kv_\perp^2+\frac{\hbar^2}
{2 \mu'_\parallel} k_z \Big\} \phi'_{\la}(\kv)
\nonumber \\
& - \frac{e^2}{4\pi^3\varepsilon_0\epsilon'} \int d^3k'
\frac{\phi'_{\la}(\kv')}{(\kv-\kv')^2} = E_{\la}\phi'_{\la}(\kv),
\end{align}
where $\mu'_\parallel=\mu_\parallel \epsilon_\parallel/ \epsilon_\perp$,
$\epsilon'=\sqrt{\epsilon_\perp \epsilon_\parallel}$, and
$\phi'_{\la}(\kv) 
= \phi_{\la}(\kv_\perp, \sqrt{\epsilon_\perp/\epsilon_\parallel}k_z)$.

To solve Eq.~(\ref{IWET}) efficiently, we expand the wave functions
$\phi'_\la(\kv)$ using spherical harmonics $Y_l^m(\theta,\varphi)$
\begin{equation}
\label{WFA}
\phi'_\la(\kv) = \sum_{l,m} R_{\la,l,m}(k) Y_l^m(\theta,\varphi) \, ,
\end{equation}
where $l=0,1,2,\ldots$ and $|m|<l$ are the angular and magnetic
quantum numbers, respectively, and $R_{\la,l,m}(k)$ is the radial
component. Projecting Eq.~(\ref{IWET}) to spherical harmonics using the
ansatz~(\ref{WFA}) yields an eigenvalue problem for the radial part alone
that can be solved numerically. A detailed description of this method will
be published elsewhere \cite{PS_unpublished}. The expansion Eq.~(\ref{WFA})
converges fast in terms of angular quantum numbers. To obtain the required
wave functions, we included five $l$ states.

\end{document}